# Electron beam evolution in a successive Compton backscattering

**D.V. Gavrilenko[1]*, A.A. Savchenko[1]*, M.N. Strikhanov[1] and A.A. Tishchenko[1,2]**

[1] National Research Nuclear University "MEPhI", 31 Kashirskoe shosse, Moscow, Russia, 115409;
[2] Laboratory of Radiation Physics, Belgorod National Research University, 308015 Pobedy St. 85, Belgorod, Russia
* Corresponding author: DVGavrilenko@mephi.ru
Corresponding author: AASavchenko1@mephi.ru

**Abstract**

Inverse Compton scattering (ICS) is a unique source of highly monochromatic x-ray and gamma radiation. We investigate theoretically the cumulative effects of repeated head-on interactions between the electron beam and a train of powerful laser pulses inside a linear accelerating structure placed between mirrors of an optical resonator. We find that the longitudinal momentum spread converges exponentially to an equilibrium value due to the competition between quantum excitation (heating) and radiation friction (cooling). The predictions of the developed theory coincide very well with computer simulations. Our work establishes the necessity to account for cumulative transverse beam dynamics in the design and optimization of future stable, high-brightness ICS sources.

**Keywords:** inverse Compton scattering, x-ray radiation, Geant4, simulation

## 1. Introduction

The process of inverse Compton scattering (ICS) of laser photons by relativistic electrons is one of the most promising mechanisms for creating a bright, compact, and universal source of quasi-monochromatic radiation in the X-ray and gamma-ray bands [1-9]. Indeed, despite having significantly smaller dimensions, this source achieves radiation brightness comparable to that of a typical synchrotron's bending magnet. For many years, ICS was of interest primarily due to its astrophysical significance [10-13]. However, in the 1960s, following significant advancements in accelerator technology and the invention of the laser, it became possible to observe this phenomenon under controlled conditions in laboratories. As a result, proposals emerged for utilizing the inverse Compton effect to create gamma-ray sources applicable to nuclear physics experiments [14-17].

To date, inverse Compton scattering remains the only viable mechanism for generating an intense monoenergetic beam of gamma photons with energies ranging between 10 and 50 MeV in accelerators. Such energies are crucial for studying radioactive isotopes far removed from the stability valley, with applications spanning nuclear photonics, photo dissociation reactions, nuclear spectroscopy, etc. [18]. Another possibility to obtain such beams is using channeling radiation, but achieving such a monochromatic gamma source this way comes with challenges like crystal fragility, limited electron beam currents, and ultra-precise alignment requirements between crystals and electrons.

Despite ongoing research [19-22], practical implementations of ICS sources fall short of desired user expectations. The photon yield in existing facilities is several orders of magnitude lower than designed values required, for instance, for such an important procedure as angiography. This difference exists due to a number of unresolved issues that are crucial for creating compact and bright sources: coherent radiation generation, methods for optimizing beam collision geometry, accounting for nonlinear processes, and the influence of the scattering event on the electron beam emittance. The last of these, accounting for the impact of recoil during Compton scattering events, seems to be one of the major obstacle for creating of the high-flux ICS source exploiting many times the same electron beam interacting with a powerful laser pulses.

Here we consider the ICS source to be a facility designed as a compact 44 MeV linear accelerator with an interaction point, where the electron beam can interact with a train of powerful laser pulses, or with a single but long laser pulse. In this paper we deal with the former case and consider an infrared and green laser operating at a micro-pulse repetition rate in the tens of gigahertz. Such a design of the ICS can be used for creating fan-like monochromatic X-ray beams that are necessary for low-dose medical procedures and industrial applications to increase the irradiation speed. In Fig. 1 we schematically show the geometry of such an interaction point, where the train of the laser pulses (green circles) is being stored in the two mirror optical cavity [23]. An accelerating structure is co-axially integrated between the optical cavity mirrors to provide in-situ energy compensation for the radiative losses of the electron beam. Once the stored intracavity laser train reaches its steady-state operating parameters, a single electron bunch is injected into the system, sequentially colliding with the laser micro-pulses and generating forward-directed x-ray radiation.

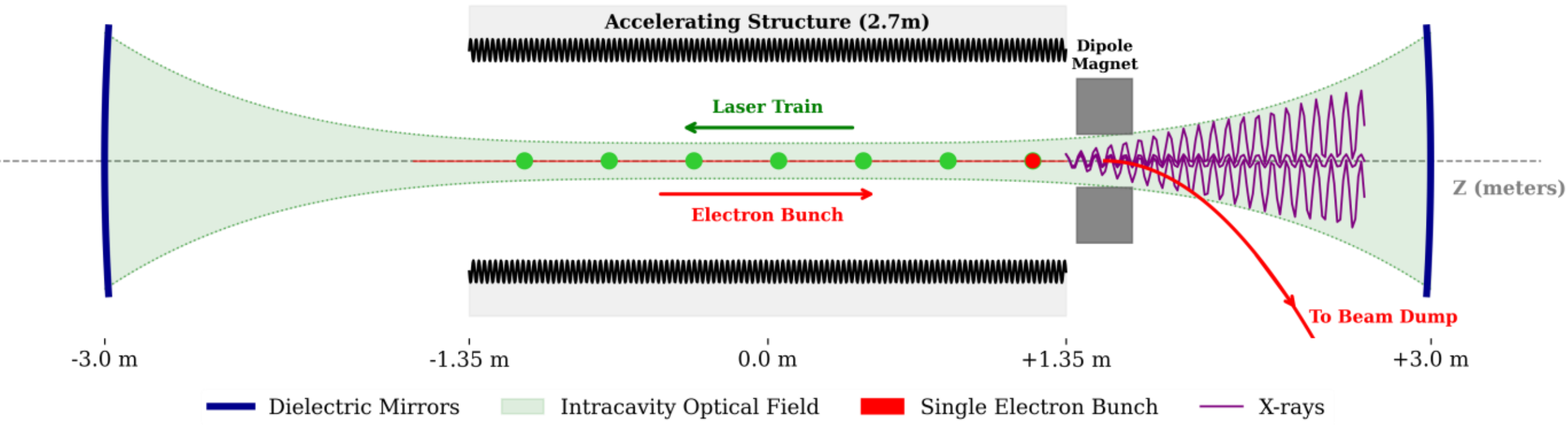

Fig. 1: Schematic representation of interactions of the single electron bunch with the laser train stored in two mirrors optical resonator inside the accelerating structure compensating energy losses of electrons. The schematic dimensions are for illustrative purposes only.

Note the driving frequency of the accelerating structure must be precisely locked to ensure that the electron bunch receives an identical longitudinal momentum kick after each successive collision. For both the infrared and green laser, the individual radiative energy losses of electrons are relatively small and their comprehensive compensation requires an accelerating structure with maximum voltage amplitude of about 100 kV on its axis. However, at sub-THz frequencies conventional closed RF cavities are highly impractical due to stringent aperture constraints. To circumvent this limitation, an oversized corrugated waveguide architecture such as, for example, high-power relativistic surface-wave generators and periodic structures operating on slow surface waves [24-26] can be utilized. As such structures are not conventional for typical accelerator here we analyze and compare two distinct operational regimes: interaction with energy compensation and interaction completely without compensation.

The majority of existing models describing ICS sources are dedicated mainly to output radiation characteristics [7, 27-29]. Recent theoretical and numerical studies, however, have increasingly focused on the recoil imparted to the electron, a phenomenon known as radiation reaction or radiation damping. In the classical regime [30-31] this is widely described by the Landau-Lifshitz equation, which treats the recoil as a continuous "cooling" force that systematically reduces the beam's normalized emittance. However, at higher electron and laser energies, a quantum treatment is necessary [31-32]. This introduces two critical factors: (1) the overall suppression of the classical radiated power, and (2) the inherent stochasticity of discrete photon emission. This stochasticity acts as a "heating" mechanism, known as "quantum excitation" (QE), which causes diffusion that increases the beam's transverse emittance and energy spread. Consequently, much of the research in this area, pioneered in works by Telnov [27], Huang and Ruth [28] and Kolchuzhkin, Potylitsyn, Strokov [29], focuses on modeling the competition between this continuous damping (cooling) and stochastic excitation (heating). Modern computational approaches employ methods ranging from lattice simulations of the entire accelerator [30-31,33] to adjoint kinetic equations to track the evolution of the beam's distribution moments (e.g., mean and variance) combined with partly quantum, partly classical Monte Carlo (MC) algorithms [29]. These MC schemes, which are now widely integrated into strong-field QED simulation codes, simulate the process as classical electron propagation interrupted by discrete, probabilistic emission events sampled from the quantum synchrotron spectrum.

Nevertheless, the classic approach to modeling proposed in above discussed works is primarily based on PIC (Particle-in-Cell) simulation [34]. This approach provides reliable results, but it is not as convenient as modeling in Geant4 [45-46], since the latter makes it possible to combine the simulation of several physical processes, such as the influence of inverse Compton scattering on beam properties, interactions with the magnetic system, and radiation of different natures. Therefore, in this article, we will focus on simulations this effect in Geant4 and on validating it by meant of the developed theoretical description. The goal of our model is to accurately predict the beam's phase-space evolution and determine the final equilibrium emittance and energy spread at which these two opposing effects balance.

In this paper, we perform theoretical investigations on electron beam evolution and extend our Geant4 application [43] by introducing novel functionalities. Based upon the theoretical results [44], we implement capabilities to simulate multi-interaction ICS sources. Our findings indicate that iterative interactions significantly influence beam characteristics, underscoring the necessity of accounting for such effects for designing the next generation of ultra-bright ICS sources.

## 2. Materials and Methods

*Theory: longitudinal momentum spread at head-on collision*

In this section, we evaluate the influence of two competing factors affecting the longitudinal momentum spread in a relativistic electron beam during its interaction with a plane laser wave during head-on collision. The first factor is broadening of the electron beam in momentum space due to the uncertainty in momentum transfer from the scattered photon. The second is that the average momentum transferred to the photon grows quadratically as the Lorentz factor increases, which causes faster electrons to decelerate more strongly. This negative feedback effectively reduces the beam spread. Therefore, we expect the momentum spread width to approach a limiting equilibrium value.

For the proper accounting for the both factors, it is necessary to know average recoil of an electron within one pass, and the variance of that recoil as well. Both can be estimated as follows:

$$\langle n\rangle = \frac{\sigma_T l\, E_0^2/4\pi}{\hbar\omega_0} = \frac{4\pi}{3}\alpha a_0^2 \frac{ct_{laser}}{\lambda_0}. \quad (1)$$

Here, $\sigma_T$ is Thomson cross-section, $l = ct_{laser}$ is duration of a laser pulse, $\lambda_0$ is its wavelength, $\hbar\omega_0$ is energy of a single photon, $E_0^2/4\pi$ is laser field energy density, $\alpha$ is fine structure constant and $a_0$ is the laser parameter. Thus, the average transferred momentum $\langle Q_z(p_z)\rangle$ of one electron with longitudinal momentum $p_z$ during one pass through the laser pulse is

$$\langle Q_z(p_z)\rangle = C\tilde{p}_z^2, \quad C = \langle n\rangle\langle q_z\rangle / \langle \tilde{p}_z^2\rangle. \quad (2)$$

Here, $q_z$ is the average transferred momentum of one electron after the interaction with an individual photon, so $Q_z = \sum_{i=1}^{n} q_{zi}$. We also use the approximation of a large number of oscillations and introduce the dimensionless momentum $\tilde{\mathbf{p}} = \mathbf{p}/mc = \gamma\boldsymbol{\beta}$, $\beta_z < 0$. We deliberately do not reduce the dependence $\langle Q_z(p_z)\rangle$ on momentum $\tilde{p}_z^2$ to the dependence on average momentum $\langle \tilde{p}_z^2\rangle$, since it defines the negative feedback feature. The variance of transferred momentum $\sigma^2(\Delta\tilde{p}_z)$ can be obtained as following. In the electron rest frame, $q_z^* = (1 - \cos\theta^*)\omega^*$ where $\theta^*$ is the polar angle of the X-ray. With the $\cos\theta^*$ distribution being $(3/8)(1+\cos^2\theta^*)$, one has $\langle q_z^*\rangle = \omega^*$ and its variance $\sigma^2(q_z^*) = (2/5)\langle q_z^*\rangle^2$. The last relation is conserved by the Lorentz transformation to the laboratory frame: $\sigma^2(q_z) = (2/5)\langle q_z\rangle^2$. Let us take into account the fluctuations of the number $n$ of Compton events. Using $Q_z = \sum_{i=1}^{n} q_{zi}$ with $q_{zi} = (1-\cos\theta_i^*)\langle q_z\rangle$ one can obtain

$$Q_z = \left(n - \sum_{i=1}^{n}\cos\theta_i^*\right)\langle q_z\rangle \quad (3)$$

and

$$Q_z^2 = \left(n^2 - 2n\sum_i \cos\theta_i^* + 2\sum_i\sum_{j<i}\cos\theta_i^*\cos\theta_j^* + \sum_i \cos^2\theta_i^*\right)\langle q_z\rangle^2. \quad (4)$$

At fixed $n$, but integrating over the scattering angles with the distribution $(3/8)(1+\cos^2\theta^*)$, one gets $\langle Q_z\rangle(n) = n\langle q_z\rangle$ and $\langle Q_z^2\rangle(n) = (n^2 + 2n/5)\langle q_z^2\rangle$. Let us assume that $n$ follows a Poisson distribution $P(n) = \exp(-\langle n\rangle)\langle n\rangle^n/n!$. Summing over $n$ the expressions for $\langle Q_z(n)\rangle$ and $\langle Q_z^2(n)\rangle$, weighted by $P(n)$, we get

$$\langle Q_z\rangle = \langle n\rangle\langle q_z\rangle, \quad \langle Q_z^2\rangle = \left(\langle n^2\rangle + 2\langle n\rangle/5\right)\langle q_z\rangle^2. \quad (5)$$

Taking into account the property $\langle n^2\rangle = \langle n\rangle^2 + \langle n\rangle$ of the Poisson distribution,

$$\sigma^2(Q_z) \equiv \langle Q_z^2\rangle - \langle Q_z\rangle^2 = (7/5)\langle n\rangle\langle q_z\rangle^2. \quad (6)$$

The longitudinal momentum transfer fluctuates due to its decomposition in discrete recoils from Compton scatterings around its' mean value $\langle Q_z\rangle$ with its' variance $\sigma^2\left(\Delta\tilde{p}_z\right)$. Thus, the final longitudinal momentum distribution can be calculated from the initial distribution as follows

$$\frac{dN'\left(\tilde{p}'_z\right)}{d\tilde{p}'_z}=\int d\tilde{p}_z\left|\frac{d\tilde{p}_z}{d\tilde{p}'_z}\right|\frac{1}{\sqrt{2\pi}\sigma\left(Q_z\right)}\times$$
$$\times\exp\left(-\frac{1}{2}\frac{\left(\left(\tilde{p}'_z-\tilde{p}_z\right)-C\tilde{p}_z^2\right)^2}{\sigma^2\left(Q_z\right)}\right)\frac{1}{\sqrt{2\pi}\sigma\left(\tilde{p}_z\right)}\exp\left(-\frac{1}{2}\frac{\left(\tilde{p}_z-\left\langle Q_z\right\rangle\right)^2}{\sigma^2\left(\tilde{p}_z\right)}\right), \tag{7}$$

where $\mathbf{p}$ and $\mathbf{p}'$ are the momenta before and after the light target.

Under the integral, there are two factors: the first accounts for the uncertainty in the momentum transfer with variance $\sigma^2\left(Q_z\right)$, and the second for the initial Gaussian distribution with variance $\sigma^2\left(\tilde{p}_z\right)$. For x-ray range, we can assume that Compton scattering changes the momentum of relativistic electrons insignificantly, so the Jacobian's deviation from unity is negligible $\left|\left|\frac{d\tilde{p}_z}{d\tilde{p}'_z}\right|-1\right|\ll 1$, and the Jacobian itself can be approximated as 1. In the gamma range this statement may not be true, so one then needs to take the Jacobian into account.

The integration in Eq. (7) is not a simple convolution, since $Q_z$ depends on $\tilde{p}_z$. But it can be made analytically in two cases. First, let us consider the case where the width of the initial distribution is much smaller than the momentum transfer uncertainty: $\sigma\left(\tilde{p}_z\right)\ll\sigma\left(Q_z\right)$. In this case, the final distribution has the form

$$\frac{dN'\left(\tilde{p}'_z\right)}{d\tilde{p}'_z}\propto\exp\left(-\frac{1}{2}\frac{\left(\tilde{p}'_z-\left\langle\tilde{p}_z\right\rangle-\left\langle\Delta\tilde{p}_z\right\rangle\right)^2}{\sigma^2\left(Q_z\right)}\right) \tag{8}$$

with the width $\sigma\left(Q_z\right)$. The result can also be obtained classically. Actually, accounting for the initial fluctuation of the Lorentz factor $\delta\gamma=\gamma-\left\langle\gamma\right\rangle$ in the formula $\gamma'=\gamma-C\gamma^2$, we can write

$$\gamma'=\left(1-C\left\langle\gamma\right\rangle\right)\left\langle\gamma\right\rangle+\left(1-2C\left\langle\gamma\right\rangle\right)\delta\gamma+C\left(\delta\gamma\right)^2. \tag{9}$$

Neglecting the last term, we obtain the same variance as in Eq. (8). Thus, the momentum transfer uncertainty provides a lower limit for the longitudinal momentum spread.

In the second case, the momentum transfer uncertainty is much smaller than the width of the current longitudinal momentum distribution $\sigma\left(Q_z\right)\ll\sigma\left(\tilde{p}_z\right)$. Upon integration, we can obtain the final distribution in the form

$$\frac{dN'}{dp'_z}\propto\exp\left(-\frac{1}{2}\frac{\left(\tilde{p}'_z-\left\langle\tilde{p}_z\right\rangle\right)^2}{\sigma^2\left(\tilde{p}_z\right)\left(1+2C\left\langle\tilde{p}_z\right\rangle\right)^2}\right). \tag{10}$$

The final square width for this case is $\sigma^2\left(\tilde{p}'_z\right)=\sigma^2\left(\tilde{p}_z\right)\left(1+2C\left\langle\tilde{p}_z\right\rangle\right)^2$. Here, we consider, that electrons are re-accelerated after each pass to compensate the average energy loss $\left\langle\Delta\tilde{p}_z\right\rangle$. Considering that $\tilde{p}_z<0$, we see that in this case, the spread decreases indefinitely. The result is expected, since we are ignoring the momentum transfer uncertainty. However, knowing that a particular certain lower limit exists, the following can be written

$$\sigma^2\left(\tilde{p}_{k,z}\right)=\rho\sigma^2\left(\tilde{p}_{k-1,z}\right)+\sigma^2_{corr}\left(\tilde{p}_z\right), \tag{11}$$

where $\rho=\left(1-2C\gamma\right)^2$, $\left\langle\tilde{p}_z\right\rangle=\gamma\beta_z\simeq-\gamma$ and $\sigma^2_{corr}\left(\tilde{p}_z\right)$ is a correcting term.

Consequently, we obtain a recurrent formula for the width of the longitudinal momentum distribution. It is then straightforward to write an expression for the width after the $k$-th interaction as a function of the initial width in the form of a partial sum of a series. Calculating this sum and returning to $\langle \tilde{p}_z \rangle = \gamma\beta_z \simeq -\gamma$, we have

$$\sigma^2\left(\tilde{p}_{k,z}\right) = \sigma^2_{corr}\left(\tilde{p}_z\right)\frac{1-\rho^k}{1-\rho} + \rho^k \sigma\left(\tilde{p}_z\right). \quad (12)$$

Since the momentum change due to radiation friction is small compared to the initial momentum, namely $C\gamma \ll 1$, we can use the following limiting relation for Euler's number:

$$\left(1+\xi^{-1}\right)^{\xi} = e, \quad \xi = -\left(2C\gamma\right)^{-1}, \ |\xi| \gg 1. \quad (13)$$

Therefore, the Eq. (12) reads

$$\sigma^2\left(\tilde{p}_{k,z}\right) = \frac{\sigma^2_{corr}\left(\tilde{p}_z\right)}{4C\gamma}\left(1-\exp\left(-4kC\gamma\right)\right) + \exp\left(-4kC\gamma\right)\sigma^2\left(\tilde{p}_z\right).$$

(14)

Thus, the spread converges exponentially to its equilibrium value

$$\sigma^2_{eq}\left(\tilde{p}_z\right) = \frac{\sigma^2_{corr}\left(\tilde{p}_z\right)}{4C\gamma}. \quad (15)$$

For the proper estimation of this equilibrium value, one has to calculate the integral in Eq. (7) exactly, which is not possible analytically. Nonetheless, we can sequentially calculate the integrals numerically for each beam-laser interaction. Each time, we find the resulting momentum distribution width and substitute it into the next integral. For the case without energy restoring, we recalculate $\langle \tilde{p}_z \rangle$ at each step. The results are shown in Figs. 3 and 5 and discussed below.

*Simulation*

Another way to look at the above considered effect is to carry out simulation via our Geant4-based code [43], which from the point of view of the ISC radiation generation was benchmarked with a help of the well-known codes such as CAIN [45] and code for ISC primary photon generation in Geant4 [46]. In our code a laser beam is substituted by a fixed target with a uniform distribution of a laser parameter $a_0$. We call this object a light target with a width equal to double laser waist $w_0$ and length equal to $l/\left(1+\beta_z\right)$. In our simulation and numerical calculation, we consider the light target with following properties: $t_{laser} = 15\,ps$, $l/\left(1+\beta_z\right) \approx 2.25\,mm$, $w_0 = 150\,\mu m$, $\lambda_0 = 530\,nm\,or\,1060\,nm$, $a_0 = 0.0381$. The laser train consists of 600 pulses with a spacing between each pulse equal to $t_{laser}$. To provide the specified laser train parameters using high-gain optical cavity, a standard commercial picosecond laser with a modest average power of 1–3 W and a micro-pulse energy of $\sim 20\,\mu J$ is sufficient.

The electron beam interacting with the target has mean kinetic energy of 44 MeV ($\gamma \approx 87$). Its initial transverse radius (HWHM) at the entrance to the light target is set to approximately $100\,\mu m$, ensuring it remains safely embedded within the laser profile to maximize the interaction volume and suppress space-charge expansion. The initial parameters in 6D phase space take the standard form: $\left(x_0, y_0, p_{x0}/p_0, p_{y0}/p_0, p_{z0}/p_0, W_0, z_0\right)$, where $x_0, y_0$ are electron beam transverse coordinates, $p_{x0}/p_0, p_{y0}/p_0, p_{z0}/p_0$ are its momentum in the form of unit vector, and $W_0$ is the beam energy. The coordinate $z_0$ represents the longitudinal position relative to the micro-pulse center and is treated as a constant value equal to $2.25\,mm$. The electron beam normalized RMS emittance is chosen to be $0.3\,mm\,mrad$. The total interaction length of the electron beam and the train of laser pulses is $2.7\,m$. In this case, the focal beta-function value of $2.87\,m$ allows the electron beam radius to be preserved practically constant. Beam population in our simulation is $10^5$, while the physical constraints for the chosen lattice permit a bunch population up to $10^9$ electrons.

To simulate multiple interactions we consider our ICS source as linear one consisting of multiple interaction points and follow the algorithm: the electron beam i) is generated being centered $2.25\,mm$ before the middle of the light target, ii) interacts with the target; if ICS photon is produced, then its momentum is subtracted from the momentum of the incident electron; finally, when electron passes another $2.25\,mm$ after the middle of the light target, the program records new parameters of the electron and returns them via creating new input file to repeats the simulation. Because during each simulation electrons generate x-ray photons and lose considerable amount of energy, their energy must be somehow restored. In our code, we are trying to simulate work of the radio frequency cavity to its simplest extent via conserving the mean energy of the electron beam. To do that we retrieve mean values of the longitudinal momentum before and after interaction, find the difference between them, and add obtained value to each electron longitudinal momentum renormalizing its kinetic energy as well. Addition of the same value of longitudinal momentum to each particle of the beam is valid if one considers electron bunches shorter than at least a quarter of the cavity wavelength. For the case without energy restoring we skip this step in the algorithm.

## 3. Results

Let us evaluate the effect of multiple interactions of the electron beam with the laser train. In Fig. 2, we show simulated distribution of $p_z c$, which describes the longitudinal momenta for the electrons before (blue), after 200 (cyan), 400 (magenta), and 600 (red) interactions of the beam with different initial energy spreads with the green and infrared laser. For the infrared laser here we show results for the only 0% energy spread as other distributions do not demonstrate pronounced difference between interactions. Fig. 3 shows standard deviation value of distributions from Fig. 2 for each pass.

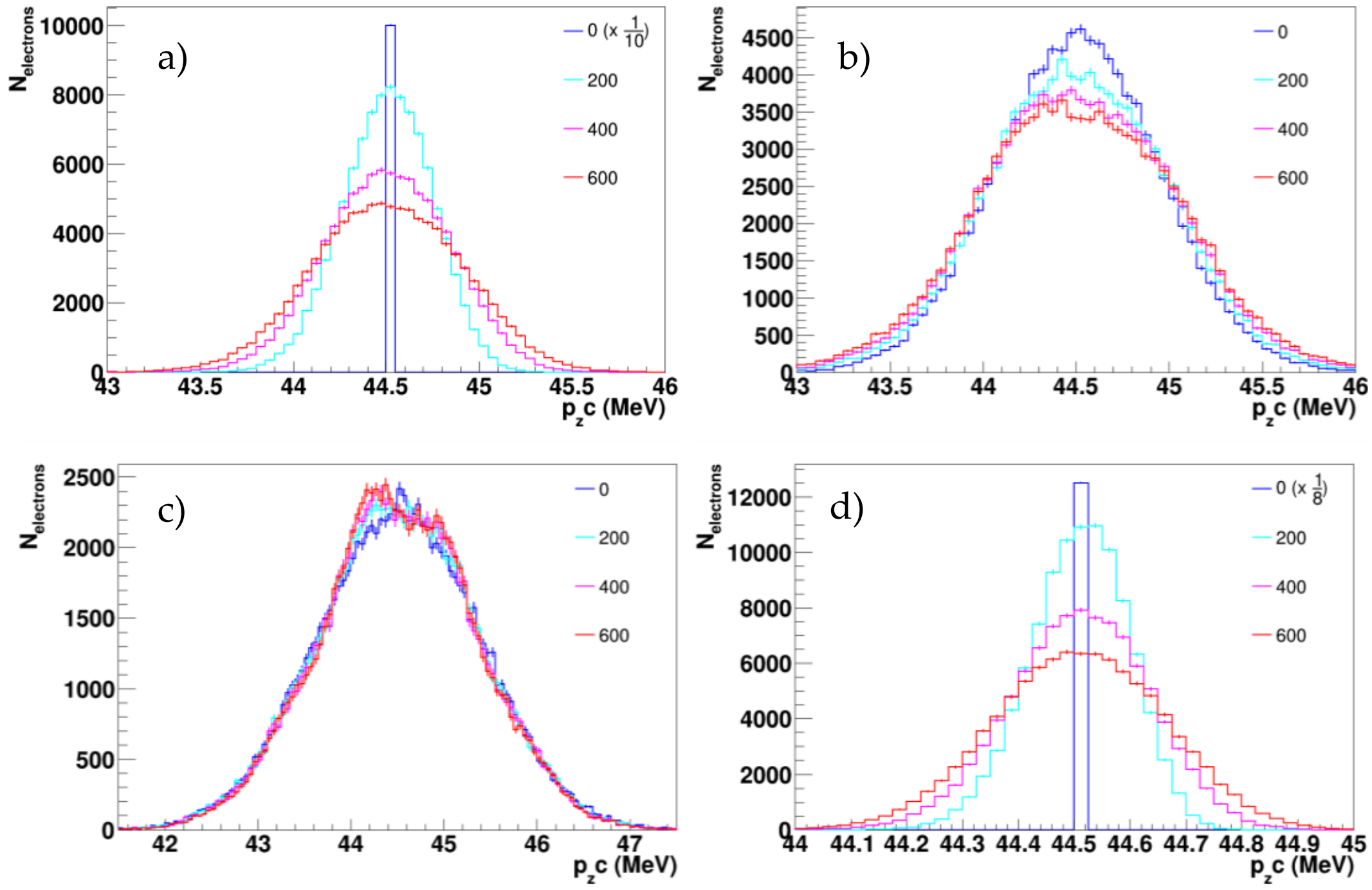

Fig. 2: Simulated evolution of the longitudinal momentum distribution of the electron beam with different initial energy spread: a, d – 0%, b – 1%, c – 2%, before interaction with laser (blue), after 200 (cyan), 400 (magenta), and 600 (red) passes. a,b,c) – green laser, d) – infrared laser.

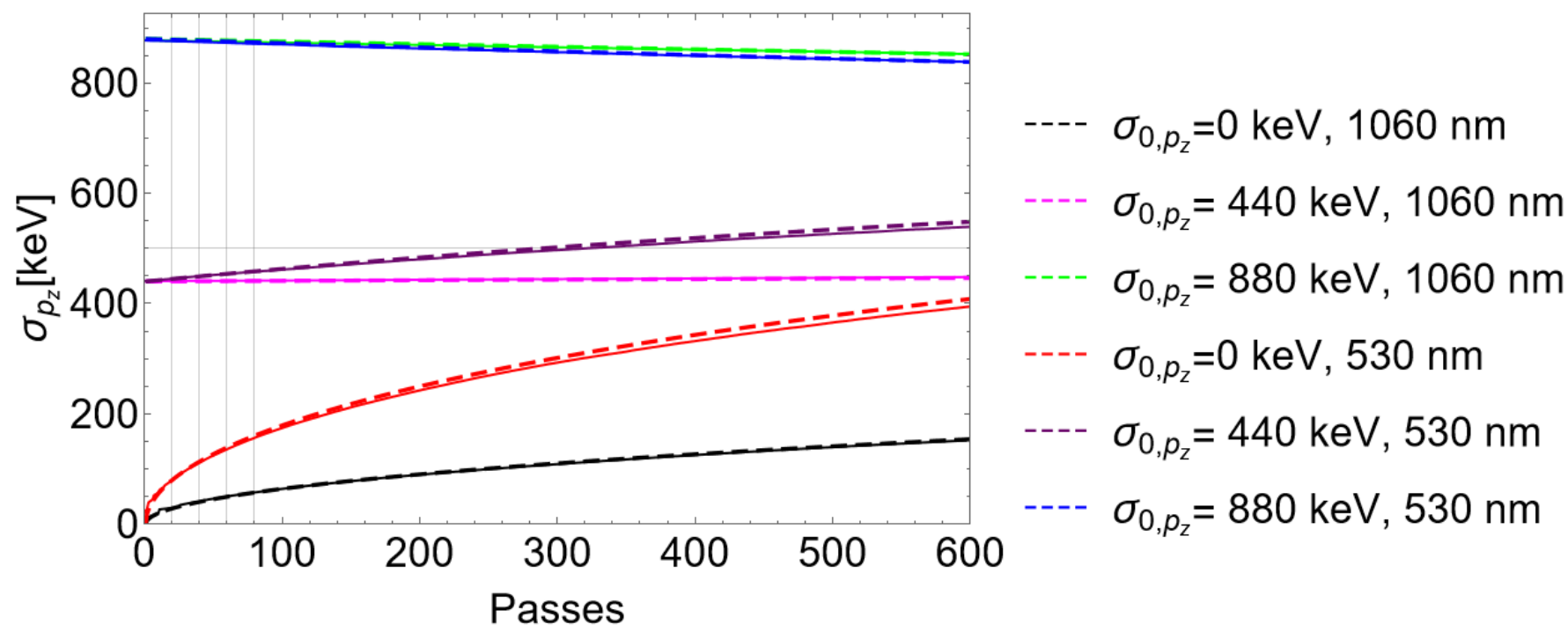

Fig. 3: Evolution of the longitudinal momentum ( $p_z c$ ) dispersion of the electron beam with different initial energy spread: black and red curves – 0%, magenta and dark purple curves – 1%, green and blue curves – 2%. The black, magenta and green curves are for the infrared laser. The red, dark purple and blue curves are for the green laser. Solid lines are from simulation and dashed lines were obtained with sequentially numerical integration of Eq. (7).

In Figs. 4a,b, we show the effect of multiple interactions of the electron beam 0% energy spread with the green and infrared laser in the case when the energy losses are not compensated. In these figures, one can see not only broadening of the distributions but also a drastically changed mean energy of the beam. Because the evolution of the longitudinal momentum dispersion without energy compensation appeared to be almost the same as one shown in Fig. 3, we do not show it here.

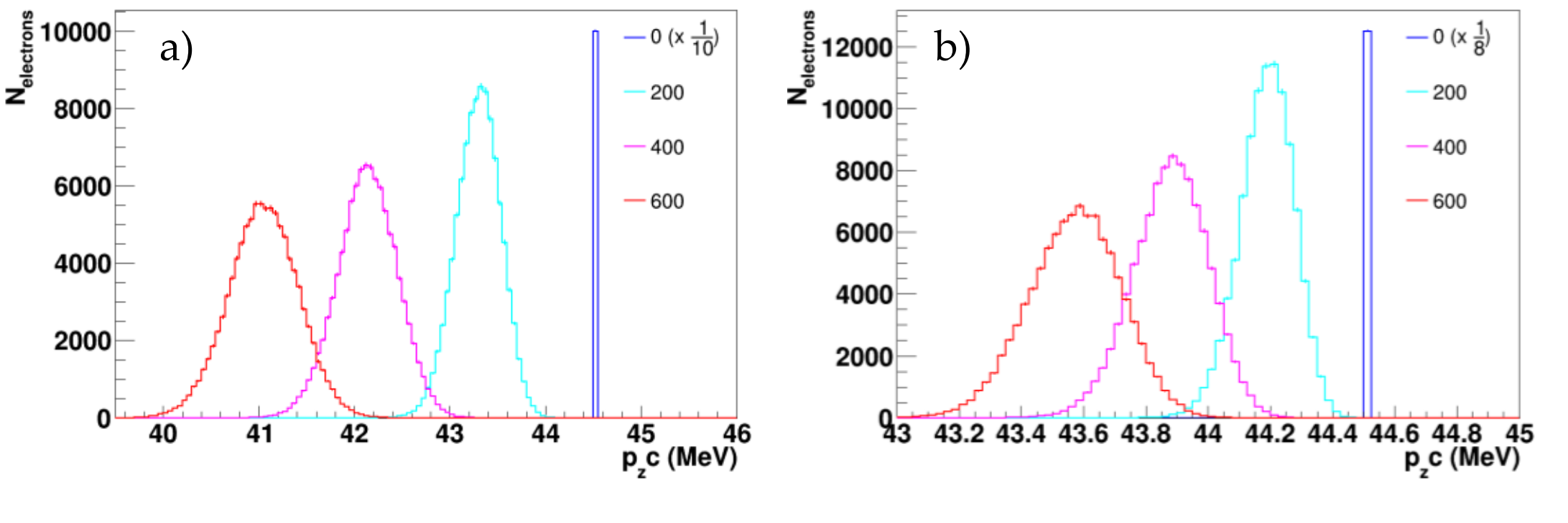

Fig. 4: Simulated evolution of the longitudinal momentum distribution of the electron beam with 0% initial energy spread before (blue), after 200 (cyan), 400 (magenta), and 600 (red) interactions with a) green and b) infrared laser.

## 4. Discussion

From simulation results and numerical integration of analytical formula as well as in Fig. 3 it is observed that an initially monochromatic beam rapidly acquires substantial energy dispersion. The beam with 1% initial spread slowly changes, and beam with a 2% spread, conversely, becomes narrower in terms of its longitudinal momentum distribution. In addition, it is evident from Fig. 3 that all distributions of beam longitudinal momentum tend to an equilibrium value. These simulation results are in a good agreement with the theoretical formulas of Sec. 2, as well as with other predictions from the field of beam damping. The slight difference in the speed of convergence for the green laser as causing higher x-ray energies than infrared one could arise due to used assumptions, such as the assumption of a Poisson distribution of Compton scattering events, a simplified energy recovery mechanism in which the center of the momentum distribution shifts to its original position after each interaction with the laser. The Fig. 3 also shows that changing the laser wavelength causer shifting the equilibrium values for the longitudinal momentum.

It is interesting to note that the restoring of the mean energy has practically no effect on the evolution of the longitudinal momentum variance. Across 600 successive collisions, the curves obtained with the energy recovery procedure enabled and those without it coincide almost completely. This phenomenon is particularly remarkable because, as demonstrated in Figure 4, although the electron loses a significant fraction of its energy, this degradation does not noticeably impact its longitudinal momentum spread. Concurrently, it is important to understand that the distributions displayed in Figure 4 indicate a substantial drop in electron energy, which will inevitably lead to a broadening of the spectral characteristics of the Compton radiation. Yet, the total number of emitted photons does not decrease significantly with this energy decrease. This is manifested by the minimal changes in the variance of the longitudinal momentum dispersion evolution. Consequently, for specific applications, in which a narrow spectral linewidth is not a strict requirement, such a radiation source can be effectively used even without implementing an accelerating cavity.

Thus, in this study through theoretical analysis and rigorous simulations, we provide key insights into the evolution of electron beam characteristics during repeated interactions with laser pulses. These interactions profoundly impact essential beam attributes, such as energy spread and divergence, necessitating careful consideration in the design of high-brightness ICS sources.

Our theoretical estimates and numerical simulations indicate that an equilibrium value exists for the longitudinal momentum spread of the electron beam. This result is expected, as the radiation friction force provides negative feedback, effectively reducing the volume in momentum space. This effect enables the application of laser cooling technology and underlies the principle of damping rings.

These findings regarding the equilibrium state have a profound practical implication on the total photon yield and source brightness. Based on our numerical estimations, a single head-on collision with the infrared laser micro-pulse yields an average number of photons of 0.092 per individual electron. The cumulative yield scales linearly to 55.2 photons per single electron across the 600 successive interactions. For the electron bunch population of $10^9$, this provides a high-intensity output of $5.52\times10^{10}$ photons per single macro-pulse. At an advanced macro-repetition rate of 200 Hz to 300 Hz, the system delivers an average photon flux of $1.10\times10^{13}$ to $1.66\times10^{13}$ photons/s. For the green laser the photon yield is approximately twice as high.

These findings underscore the importance of using realistic laser pulse parameters and optimization of the procedure of integrated beam dynamics simulations into the design process for advanced ICS sources. Future efforts should focus on refining models to fully capture these complexities, ensuring optimal performance and stability in practical applications.

**Author Contributions:** Conceptualization, A.A.T., D.V.G and A.A.S.; methodology, D.V.G and A.A.S.; software, A.A.S.; validation, A.A.T, D.V.G. and A.A.S; formal analysis, D.V.G. and A.A.S.; investigation, D.V.G. and A.A.S.; resources, A.A.T. and M.N.S.; writing—original draft

preparation, D.V.G. and A.A.S.; writing—review and editing, A.A.T and M.N.S; visualization, D.V.G. and A.A.S.; supervision, A.A.T. and M.N.S.; project administration, A.A.S.; funding acquisition, A.A.S., A.A.T and M.N.S. All authors have read and agreed to the published version of the manuscript.

**Funding:** This research was funded by the Ministry of Science and Higher Education of the Russian Federation under the strategic academic leadership program «Priority 2030».

**Data Availability Statement:** The original data presented in the study are available on reasonable request from the corresponding author.

**Acknowledgments:** We are grateful to P.G. Shapovalov for fruitful discussions.

**Conflicts of Interest:** The authors declare no conflicts of interest.

## Abbreviations

The following abbreviations are used in this manuscript:

| | |
|---|---|
| ICS | Inverse Compton scattering |
| HWHM | Half width half maximum |